\newif\ifreport\reporttrue
\documentclass[conference]{IEEEtran}
\usepackage{amsfonts}
\usepackage{amsthm}
\usepackage{bm}
\usepackage{amsmath,amssymb}
\usepackage{graphicx}
\usepackage{url}
\usepackage{cite}
\usepackage{hyperref,stfloats,theorem}
\usepackage[tight,footnotesize]{subfigure}

\graphicspath{{./Figures//}}
\usepackage[normalem]{ulem}
\usepackage{color}
\usepackage{verbatim}
\usepackage{breakurl}
\usepackage{caption}
\begin{document}

\title{Life-Add: {Life}time {Ad}justable {D}esign for WiFi Networks with Heterogeneous Energy Supplies}

\author{\IEEEauthorblockN{ Shengbo Chen\IEEEauthorrefmark{1}\IEEEauthorrefmark{4}, Tarun Bansal\IEEEauthorrefmark{2}\IEEEauthorrefmark{4},
Yin Sun\IEEEauthorrefmark{1}\IEEEauthorrefmark{4},  Prasun Sinha\IEEEauthorrefmark{2}
and  Ness B. Shroff\IEEEauthorrefmark{3} }
\IEEEauthorblockA{\IEEEauthorrefmark{1}\IEEEauthorrefmark{3}Department of ECE, The Ohio State University}
\IEEEauthorblockA{\IEEEauthorrefmark{2}\IEEEauthorrefmark{3}Department of CSE, The Ohio State University}
{Email: \{chens,shroff\}@ece.osu.edu,
\{bansal,prasun\}@cse.ohio-state.edu, sunyin02@gmail.com}
\IEEEauthorblockA{\IEEEauthorrefmark{4}\textbf{Co-primary authors}}
}

\maketitle
\begin{abstract}

WiFi usage significantly reduces the battery lifetime of handheld devices such as smartphones and tablets, due to its high energy consumption. In this paper, we propose ``Life-Add'': a \uline{Life}time \uline{Ad}justable \uline{d}esign for WiFi networks, where the devices are powered by battery, electric power, and/or renewable energy. In Life-Add, a device turns off its radio to save energy when the channel is sensed to be busy, and sleeps for a random time period before sensing the channel again. Life-Add carefully controls the devices' average sleep periods to improve their throughput while satisfying their operation time requirement. It is proven that Life-Add achieves near-optimal proportional-fair utility performance for single access point (AP) scenarios. Moreover, Life-Add alleviates the near-far effect and hidden terminal problem in general multiple AP scenarios. Our ns-3 simulations show that Life-Add \emph{simultaneously} improves the lifetime, throughput, and fairness performance of WiFi networks, and coexists harmoniously with IEEE 802.11.\end{abstract}

\section{Introduction}
WiFi is one of the most successful wireless Internet access technologies, which has become prevalent for handheld mobile devices, such as smartphones and tablets. One major concern of the current IEEE 802.11 WiFi design is its high energy consumption.  
WiFi, along with GPS and Bluetooth, uses a significant amount of energy,
which causes rapid drops in battery level, especially for smartphones. In
fact, it has been reported that the IEEE 802.11 WiFi design may increase the smartphone's energy consumption by up to 14 times \cite{Zhang:2011:EEI:2030613.2030637}.

Currently, there are two categories of works that focus on prolonging the operation time of WiFi devices. The first class aims at reducing the energy consumption of WiFi radios, e.g., \cite{Zhang:2011:EEI:2030613.2030637,DBLP:conf/mobisys/RoznerNRR10,Baiamonte:2006:SED:1147589.1147605}. The second class uses a renewable solar charger or a portable battery to provide mobile energy replenishment without being confined to the immovable electric power sources on the wall, e.g., \cite{SolarMobileCharger1,portable_batteries}.
Beyond these, it would be of great interest to design lifetime-tunable WiFi devices where the throughput performance can be also improved.



In this paper, we propose \emph{Life-Add}: a \underline{Life}time \underline{Ad}justable \underline{d}esign for WiFi networks, where the devices are powered by battery, electric power, and/or renewable solar energy. In this design, a device senses the channel availability when it has a packet to send. If the channel is
sensed to be busy, it goes to sleep for an exponentially distributed random  period of time and then wakes up to sense the channel availability again. If the channel is idle, it transmits the packet. Life-Add carefully controls the average sleep periods of the WiFi devices to improve their throughput and fairness performance, while satisfying their operation time requirement. We first formulate a proportional-fair network utility maximization problem for  asynchronous WiFi networks in  single AP scenarios. Unlike traditional synchronized wireless networks, such as cellular networks, the channel access probabilities of the WiFi devices are coupled due to channel contentions. As a result, the attained network utility maximization problem turns out to be a difficult non-convex optimization problem. However, we manage to develop a low-complexity solution and prove that it is near-optimal to the
problem. In general multiple AP scenarios, due to the complicated  interference environment, the WiFi networks suffer from the well-known ``\textit{near-far effect}'' and the ``\textit{hidden terminal
problem}''. Life-Add introduces device collaboration to alleviate the near-far effect and leverages congestion control to handle the hidden terminal problem. 

The contributions of this paper are summarized as follows:
\begin{itemize}
\item We propose a new model for the WiFi transmission behaviors, and characterize the throughput and energy consumption performance of each device. Thanks to the memoryless property of the exponential sleep period distribution, the attained performance metric functions are quite simple. This forms the basis for the development of an optimization framework, which leads to our control strategy.

\item We formulate a non-convex proportional-fair utility maximization problem for the single AP scenarios, and develop a simple solution, i.e., Life-Add, to control the average sleep time periods. We prove that Life-Add is near-optimal, if the carrier sensing time is negligible compared to the packet-plus-ACK transmission time. This condition is satisfied for practical WiFi packet transmission parameters; see the Remark in Section \ref{sec:performance_ana} for details.

\item We generalize Life-Add to deal with the near-far effect and the hidden terminal issue heuristically. Our key contribution here is to develop a new device collaboration scheme, which  improves the throughput and fairness performance. In addition, the Life-Add design is simple and easy to implement.


\item Our ns-3 simulations show that Life-Add simultaneously improves the lifetime, throughput, and fairness performance of WiFi networks.  A brief performance comparison between Life-Add and IEEE 802.11b is provided in Table \ref{tab:compare} for the simulation scenario of Fig. \ref{fig:mc:hetro:utiltput-varyingNoDevices}.
  We also show that Life-Add can coexist harmoniously with   IEEE 802.11: if some devices upgrade from IEEE 802.11 to Life-Add, the throughput performance
improves for both the devices switching to Life-Add and those sticking to IEEE 802.11.
\end{itemize}

Compared to IEEE
802.11, first, the lifetime improvement of Life-Add is due to turning off the radio
during the backoff period. Second, by  controlling the transmission parameters    carefully  and introducing collaboration among the devices, Life-Add has a smaller packet collision probability and
thus leads to  better throughput performance. Finally, by  providing higher priorities to the low-throughput devices, Life-Add can improve the fairness performance.

\begin{table}
\caption{Performance comparison of different WiFi designs.}\label{tab:compare}
\begin{center}
\begin{tabular}{l|c|c|c}
\hline
 & IEEE 802.11b & IEEE 802.11b   & Life-Add \\
& & with RTS-CTS & \\ \hline
Avg Lifetime& 147.87 min & 168.79 min  & 252.45 min  \\ \hline
Avg throughput & 0.9568 Mbps & 1.0338 Mbps & 1.41 Mbps \\ \hline
Jain's fairness index \cite{Jain:1991} & 0.427 & 0.402 & 0.691 \\ \hline
successful ACK & 68.3\% & 98.6\% & 87.45\%\\ \hline
\end{tabular}
\end{center}
\vspace{-15pt}
\end{table}


\section{Related Works}\label{sec_stateoftheart}
Performance analysis of carrier sensing multiple access (CSMA) schemes for WiFi networks have been widely investigated since its seminal paper \cite{840210}. However, the two dimensional Markovian chain model in \cite{840210} makes it difficult to control the WiFi transmissions. Recently, a queue based  scheduling scheme called Q-CSMA, was proven to be throughput optimal~\cite{5340575}. These studies focus on throughput enhancement, assuming the WiFi radio is turned on all the time for sensing, transmitting, or receiving, which is not energy efficient.

Energy efficient scheduling schemes have been developed for synchronized wireless networks (see the recent papers \cite{Neely:2006:EOC:2263442.2272284, Lin:2010:LDE:1816262.1816275,6195775,Huang:2011:UOS:2107502.2107531} and the references therein). 
These works focused on slotted wireless networks, such as cellular networks. Therefore, they are not applicable to asynchronous WiFi networks.

Sleep-wake scheduling schemes have been developed for wireless sensor networks to prolong the battery lifetime (see \cite{5339116,Zhao:2012:SPM:2379776.2379783,5451759} and the references therein). While these schemes apply to asynchronous wireless networks, their focus is on light traffic sensor networks where packet collisions seldom occur. However, packet collision is one of the most serious problems in WiFi networks.

Recently, energy efficient WiFi designs have received significant research attention due to the emerging handheld WiFi devices with small battery capacity, e.g., \cite{Zhang:2011:EEI:2030613.2030637,DBLP:conf/mobisys/RoznerNRR10,Baiamonte:2006:SED:1147589.1147605}. In  \cite{Zhang:2011:EEI:2030613.2030637}, the authors proposed to reduce the power consumption in idle-listening by lowering
the clock-rate in hardware.
The authors of \cite{DBLP:conf/mobisys/RoznerNRR10} found that IEEE 802.11's Power Save Mode (PSM) was not efficient in the presence of competing background traffic, and proposed a scheduling algorithm to resolve it.
In \cite{Baiamonte:2006:SED:1147589.1147605}, the devices enter the sleep mode during the backoff like our scheme, but without controlling the sleep period distribution. While these studies improve the energy efficiency of WiFi networks, they cannot improves the lifetime, throughput, and fairness performance simultaneously.

\section{System Model}\label{sec_systemmodel}
We first consider a WiFi network with 1 AP and $N$ devices, where all the devices are associated with the AP.
Extensions to general multiple AP scenario are discussed in Section \ref{sec:multiple}. 




\subsection{Energy Model}
Consider that device $n$ will be used for a continuous time duration no shorter than $T_{target,n}$ before its battery dies out, where $T_{target,n}$ is called the \emph{target lifetime} of device $n$. The lifetime constraint of device $n$ is given by
\begin{align}\label{eq:lifetime_con}
T_n \geq T_{target,n}, ~\forall n\in \mathcal{N},
\end{align}
where $T_n$ is the actual lifetime of device $n$ and $\mathcal{N}=\{1,2,\cdots,N\}$ is the set of devices.

The energy cost of each device consists of two parts: the
energy spent on the WiFi radio and the base energy consumption in powering the non-WiFi components, such as CPU, screen, etc.
Let $E_{RF,n}$ denote the average energy consumption rate of the WiFi circuit in device $n$ when turned on, and $E_{nonRF,n}$ denote the average energy consumption rate of the non-WiFi components in device $n$. We assume that, when the WiFi radio is in sleep mode, its energy consumption rate is negligible compared to $E_{RF,n}$. For example,
measurements showed that the WiFi radio in HTC Tilt mobile phone consumes an average power of 72mW in the
sleep mode while it consumes 1120 mW in the awake mode \cite{DBLP:conf/mobisys/RoznerNRR10}.
Let $B_n$ represent the amount of energy initially stored in the battery, and $r_n$  the average energy replenishment rate for device $n$ (either from renewable energy source or from electric
power supply), which can be obtained from historical data.\footnote{If a device relies only on the battery, we have $ r_n=0$.} Then, the maximum allowable energy consumption rate of the WiFi radio is
\begin{align}\label{eq:energy_rate}
e_{con,n}=\frac{B_n}{T_{target,n}}+r_n-E_{nonRF,n}, ~\forall n\in \mathcal{N}.
\end{align}

Therefore, the lifetime constraint \eqref{eq:lifetime_con} can be rewritten as the energy consumption constraint as follows
\begin{align}\label{eq:energy_con}
e_n \leq e_{con,n}, ~\forall n\in \mathcal{N},
\end{align}
where $e_n$ is the actual energy consumption rate of the WiFi radio.
In practice, the energy replenishment rate of the solar charger $r_n$ is usually low such that $r_n \leq E_{nonRF,n}$.\textit{ In order to make sure that the lifetime constraint is feasible, we require that $e_{con,n}\geq 0$. Therefore, $T_{target,n}$ is no greater than $T_{\max,n} \triangleq B_n/(E_{nonRF,n}-r_n)$, where $T_{\max,n}$ is the maximum feasible lifetime of device $n$.\footnote{If a device recharges from electric power supply, then $r_n > E_{nonRF,n}$ and $T_{\max,n}$ is infinite. }}

\begin{figure*}[!t]
\centering
\includegraphics[width=0.6\textwidth]{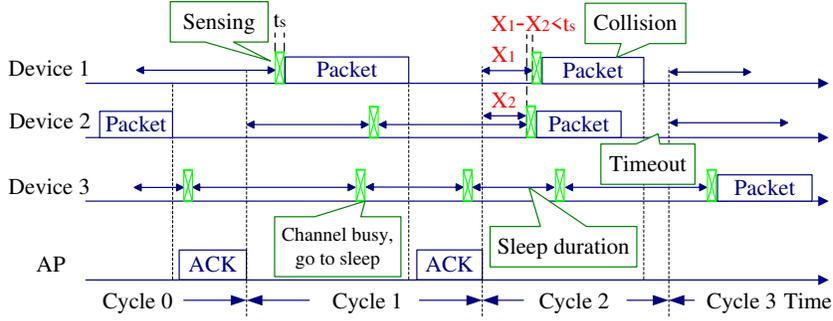}
\caption{Illustration of the sleep-wake channel contention scheme.}
\label{fig_11}
\vspace{-0.5cm}
\end{figure*}

\subsection{Sleep-Wake Channel Contention}\label{sec:sleep_wake_scheme}
We now describe our sleep-wake channel contention scheme. Consider the uplink scenario, i.e., from the devices to the AP. When a device wakes up from a sleep mode, it performs carrier sensing immediately if it has a packet to transmit.
We use $t_s$ to represent the carrier sensing time. Since $t_s$ is very
short, we assume that the energy spent on carrier sensing
is negligible; see the Remark in Section \ref{sec:performance_ana}.
If the channel is free, the device transmits the packet; otherwise,
the device goes back to sleep immediately. The sleep period for each device $n\in \mathcal{N}$ is an independent exponentially distributed random period of time with mean $1/R_n$, and the control of $R_n$ will be discussed later. 

If two devices begin transmitting within a duration shorter than the carrier sensing time $t_s$, they may not be
able to detect each other's transmissions and a collision will occur. We assume that the devices are within each other's sensing range.\footnote{ This assumption is removed in Section \ref{sec:multiple} when studying
the case with multiple AP scenarios.} Therefore, each device can  detect the transmissions of another device as long as the difference of their wake-up times is larger than $t_s$.
If the AP successfully receives the packet, it replies with an ACK. After data transmission, the WiFi radio stays awake for an ACK or a timeout. Then it goes back to sleep for another exponentially distributed time duration.
The average transmission time of a data packet is denoted by $L$ and the duration of the ACK is $t_a$. Moreover, we assume that when there is a collision, the average collision time plus timeout can be approximated to be $L+t_a$. Note that packet collisions may also occur when some devices wake up before the ACK transmission. We omit these collisions in our model to keep the formulated design problem solvable. For similar reasons, we also omit the events that some devices wake up during the timeout \cite{840210}. However, they will be considered in our ns-3 simulations in Section \ref{sec:simulation}.

We define a \emph{sleep-wake cycle} as the duration between the end
of two successive timeouts or ACKs.
Figure \ref{fig_11} illustrates the sleep-wake cycles of three contending devices.
After the first ACK, the channel becomes idle and then the one with the
smallest remaining sleep period, i.e., Device 1, wakes up and begins transmission. Since no other devices is awake in the preceding and the following period of length $t_s$, the transmission is successful, and the AP replies with an ACK.
In the next sleep-wake cycle, since Device 1 and 2 wake up and transmit within a time period shorter than $t_s$, it leads to
a collision. If the transmitting device does not receive the ACK before the timeout, it goes back to sleep for another exponentially distributed time duration and again contends for the channel to retransmit the packet.


For the downlink scenario, IEEE 802.11 defines a power saving mode (PSM), where
the AP periodically broadcasts beacons to notify which devices have pending data packets in the buffer. Each device with pending packets then sends a short packet, called ps-poll packet, to the AP for channel contention. After receiving a ps-poll packet from one winning device, the AP transmits the downlink packets to the device \cite{DBLP:journals/winet/AnastasiCGP08}.
Note that the channel contention of the downlink case is similar to the uplink case. We focus on the uplink scenario in this paper, and will extend it to the general scenario with both uplink and downlink traffic in our future work.
\section{Problem Formulation}\label{sec_problemformulation}
\subsection{Characterization of Throughput and Energy Consumption }
Let $X_n$ denote the residual sleeping period
of device $n$ when the latest sleep-wake cycle is over. Since the sleeping period of device $n$ is exponentially distributed with parameter $R_n$, $X_n$ is also exponentially distributed with parameter $R_n$, owing to the memoryless property of exponential distribution.

Let $\beta_n$ denote the probability that device $n$ obtains the channel
and transmits successfully after a sleep-wake cycle.
According to our sleep-wake channel contention scheme, device $n$ obtains the channel if $X_i\geq X_n+t_s$ for all $i\in \mathcal{N}$ and $i\neq n$.
Therefore, $\beta_n$ is given by
\begin{align}
\beta_n&= \Pr(X_i\geq X_n+t_s, \forall i\neq n) \nonumber\\
&\overset{(a)}{=} \mathbb{E}[\Pr(X_i\geq X_n+t_s, \forall i\neq n|X_n)]\nonumber\\
&\overset{(b)}{=} \mathbb{E}\left[\prod_{i\neq n}\Pr(X_i\geq X_n+t_s|X_n)\right]\nonumber\\
&=\int_{0}^{\infty}\!\left[\prod_{i\neq n}\exp(-R_i (x_n+t_s))\right]\!R_n\exp(-R_{n}x_n)dx_n\nonumber\\
&=\frac{R_n\exp(R_nt_s)}{(\sum_{i}R_i)\exp(\sum_{i}R_it_s)},\label{eq_4_1}
\end{align}
where 
 $(a)$ is due to $\Pr[A]=\mathbb{E}[\Pr(A|B)]$, and $(b)$ is due to the
fact that $X_i$ is independent for different devices.

Let $p_n$ denote the fraction of time that device $n$ transmits successfully with no collision. Each sleep-wake cycle consists of three parts: transmission/collision, ACK/timeout and idle period. The collision probability is determined by $\beta_{col}=1-\sum_n \beta_n$, because for any sleep-wake cycle, it includes either a successful transmission or a collision. 
The idle period is the shortest sleeping period among $N$ exponentially distributed variables with parameter $R_i, \forall i\in \mathcal{N}$. It is easy to show
that the idle period is exponentially distributed with parameter $\sum_i R_i$.
In our sleep-wake channel contention model in Section \ref{sec:sleep_wake_scheme}, the duration of each sleep-wake cycle is \emph{i.i.d.} By the renewal theory in stochastic processes~\cite{Gallager96}, $p_n$ is given by
\begin{align}
\label{eq_4_2}
p_n&=\frac{ \beta_n \mathbb{E}[\mathbf{trans}]}{(\sum_i\beta_i +\beta_{col}) (\mathbb{E}[\mathbf{trans}]+\mathbb{L}[\mathbf{ACK}])+\mathbb{E}[\mathbf{Idle}]}\nonumber\\
&=\frac{\beta_nL}{L+t_a+\frac{1}{\sum_{i}R_i}}\nonumber\\
&=\frac{R_n\exp(R_nt_s)}{(\frac{L+t_a}{L}\sum_i R_i+\frac{1}{L})\exp(\sum_{i}R_it_s)},
\end{align}
where $\mathbb{E}[\mathbf{trans}]$ represents the average packet transmission time $L$, $\mathbb{L}[\mathbf{ACK}]$ represents the duration of ACK $t_a$, $\mathbb{E}[\mathbf{Idle}]$ represents the
expectation of the idle duration given by ${1}/{\sum_{i}R_i}$, and we have used the assumption that
the average collision time plus timeout can be approximated to be $L+t_a$.

Let $\bar D_{n}$ be the throughput of device $n$. In practice, $\bar D_{n}$ is proportional to the fraction of successful transmission time $p_n$:
\begin{align}
\label{eq_4_0}
\bar D_{n}=p_n\alpha_n,
\end{align}
where $\alpha_{n}$ is a scaling parameter
depending on the physical layer  signal structure, channel coding and modulation. Note that  our control solution does not require knowledge of $\alpha_n$.

On the other hand, besides successful transmissions, each device $n$ may suffer from
collisions, which also costs energy. Thus, we denote  $P_n$
as the total fraction of time that the WiFi radio in device $n$ is turned on, i.e., transmitting
or waiting for the ACK. Similar to (\ref{eq_4_2}),  we can show that
\begin{align}
\label{eq_4_3}
P_n=\frac{[1-\exp(-R_nt_s)]\sum_{i}R_i+\exp(-R_nt_s)R_n}{\sum_i R_i+\frac{1}{L+t_a}}.
\end{align}
\ifreport
The derivation of \eqref{eq_4_3} is provided in Appendix \ref{App00}.
\else
The derivation of \eqref{eq_4_3} is provided in \cite{report_scheduling2013}.
\fi
Therefore, the actual energy consumption rate of the WiFi radio is $e_n=P_nE_{RF,n}$.
Hence, the energy constraints \eqref{eq:energy_con} can be rewritten as
\begin{align}
\label{eq_5_11}
P_nE_{RF,n}\leq e_{con,n}, ~\forall n\in \mathcal{N}.
\end{align}
Let us define  $b_n\triangleq  e_{con,n}/ E_{RF,n}$ as the \emph{target energy efficiency} of device $n$, then \eqref{eq_5_11} can be rewritten as
\begin{align}
\label{eq_5_1}
P_n\leq b_n, ~\forall n\in \mathcal{N}.
\end{align}
Note that this constraint is always satisfied when $b_n\geq1$.
\subsection{Problem Statement}
Suppose that each device $n$ is associated with a proportional-fair utility function $\ln(\bar D_{n})$ \cite{Kelly97chargingand}.
%
Our goal is to design the average sleep period $R_i$ for $i\in \mathcal {N}$ to maximize the total utility. This design problem can be formulated as
\begin{align}
\label{eq_5}
\max_{\vec R>0} \sum_{i\in\mathcal{N}} \ln(\bar
D_{i}),
\end{align}
subject to the per-device energy constraint \eqref{eq_5_1}.

Substituting (\ref{eq_4_2}) and (\ref{eq_4_3})
into (\ref{eq_5_1}) and (\ref{eq_5}), this problem is rewritten as the following problem:
(\textrm{Problem} \textbf{ A})
{\small\begin{align}
&\!\!\!\!\max_{R_i>0}~ \sum_{i} \ln R_i-N\ln\left(\sum_i R_i+\frac{1}{L+t_a}\right)\nonumber\\
& ~~~~~-(N-1)\sum_{i}R_it_s+N\ln\left(\frac{L}{L+t_a}\right)+\sum_{i} \ln \alpha_i\nonumber\\
&  \textrm{s.t.} ~\frac{[1\!-\!\exp(-R_nt_s)]\sum_{i}R_i\!+\!\exp(-R_nt_s)R_n}{\sum_i R_i+\frac{1}{L+t_a}}\leq b_n, \forall n.\nonumber
\end{align}}
\vspace{-3pt}

\section{Problem Solution}\label{sec_problemsolution}
\textrm{Problem} \textbf{A} is a non-convex optimization problem. We first consider a relaxed version
of \textrm{Problem} \textbf{A}, i.e.,
{\small\begin{align}
\label{eq_61}
& \max_{R_i>0}~  \sum_{i} \ln R_i-N\ln\left(\sum_i R_i+\frac{1}{L+t_a}\right)\\
& ~~~~~~~-(N-1)\sum_{i}R_it_s+N\ln\left(\frac{L}{L+t_a}\right)+\sum_{i} \ln \alpha_i\nonumber\\
&  ~~\textrm{s.t.}   R_n\leq b_n\left(\sum_i R_i+\frac{1}{L+t_a}\right), \forall n.\label{eq_6}
\end{align}}

The  only difference between the above problem and  \textrm{Problem} \textbf{A} is  the constraints.
Since $R_n\leq [1-\exp(-R_nt_s)]\sum_{i}R_i+\exp(-R_nt_s)R_n, \forall n$, the constraints of \eqref{eq_61} are looser than those of \textrm{Problem} \textbf{A}. Hence, \eqref{eq_61} serves as a performance upper bound of Problem \textbf{A}.

In the sequel, we solve \eqref{eq_61} in two cases: $\sum_i b_i\geq 1$ and $\sum_i b_i<1$, and show in Theorem 1 that the obtained solution is asymptotically optimal for the original Problem \textbf{A} as $\frac{t_s}{L+t_a}\!\rightarrow\!0$.
%
\subsection{The Case of $\sum_i b_i\geq 1$}\label{sec:sumb>1}

In this case, we provide an approximate solution to Problem \eqref{eq_61} in two steps:
\begin{enumerate}
\item[(1)] Define an auxiliary variable $y=\sum_i R_i$. Given that $y>0$, $R_i$ is determined by the following approximate problem:
{\small\begin{align}\label{eq_12}
& \max_{R_i>0} \sum_{i} \ln R_i-N\ln\left(y+\frac{1}{L+t_a}\right)\\
&~~~~~~-(N-1)yt_s+N\ln\left(\frac{L}{L+t_a}\right)+\sum_{i} \ln \alpha_i,\nonumber\\
& ~~ \textrm{s.t.}~   R_n\leq b_n y, ~\forall n\in\mathcal{N}\label{eq_41}\\
&~~~~~~\sum_i R_i=y.\label{eq_42}
\end{align}}
\ifreport
$\!\!$In Appendix \ref{App1}, we show that the optimum solution to \eqref{eq_12} is determined by
\else
$\!\!$In \cite{report_scheduling2013}, we show that the optimum solution to \eqref{eq_12} is determined by
\fi
{\small\begin{equation}
\label{eq_13}
R_n=\min \{b_n,c^*\}y,
\end{equation}}
$\!\!$where $c^*$ satisfies
{\small\begin{equation}
\label{eq_14}
\sum_i\min \{b_i,c^*\}=1.
\end{equation}}


\item[(2)]
Substituting (\ref{eq_13}) and (\ref{eq_14}) into \eqref{eq_12},
we get
{\small\begin{align}
\label{eq_15}
&\!\!\!\!\!\!\!\max_{y>0} N \ln y-N\ln\left(y+\frac{1}{L+t_a}\right)-(N-1)yt_s\nonumber\\
&+N\ln\!\left(\!\frac{L}{L\!+\!t_a}\!\right)\!+\!\sum_{i} \ln \alpha_i\!+\!\sum_i \ln[\min \{b_i,c^*\}],
\end{align}}$\!\!$
which is an one-dimensional convex optimization problem of $y$. Taking the derivative of the objective function and setting it to be zero, yields
\begin{align}
\label{eq_16}
y^*=-\frac{1}{2(L+t_a)}+\frac{\sqrt{1+\frac{4N(L+t_a)}{(N-1)t_s}}}{2(L+t_a)}.
\end{align}
Substituting $y=y^*$ into \eqref{eq_13}, the sleep period parameter $\{R_i\}$ is derived.
\end{enumerate}
Note that \textrm{Problem} \eqref{eq_12} has more restricted energy constraints than \textrm{Problem} \eqref{eq_61}. Hence
our solution \eqref{eq_13}, \eqref{eq_14}, and \eqref{eq_16} is feasible for \textrm{Problem} \eqref{eq_61}. Since \eqref{eq_13}, \eqref{eq_14}, and \eqref{eq_16} achieves an objective value given by \eqref{eq_15}, the optimum objective value of Problem \eqref{eq_61} is lower bounded by \eqref{eq_15}.

\subsection{The Case of $\sum_i b_i< 1$}\label{sec:sumb<1}
Substituting the energy constraints \eqref{eq_6} into \eqref{eq_61}, the optimum value of \eqref{eq_61} is upper bounded by
{\small\begin{align}
\label{eq_B0}
&  \max_{R_i>0}~ \sum_{i} \ln b_i -(N-1)\sum_{i}R_it_s+\sum_{i} \ln\left(\frac{L\alpha_i}{L+t_a}\right),\\
& ~~\textrm{s.t.} \hspace{0.3cm} R_n\leq b_n\left(\sum_i R_i+\frac{1}{L+t_a}\right), \forall~n.\nonumber
\end{align}}$\!\!$
\ifreport
In Appendix \ref{App0}, it is proven that the optimum solution to \eqref{eq_B0} satisfies
\else
In \cite{report_scheduling2013}, we prove that the optimum solution to \eqref{eq_B0} satisfies
\fi
{\small
\begin{align}\label{eq1}
R_n= b_n\left(\sum_i R_i+\frac{1}{L+t_a}\right),~~ \forall~n.
\end{align}}$\!\!$
Moreover, if \eqref{eq1} holds, the optimum values of \eqref{eq_61} and \eqref{eq_B0} are identical. If $\sum_i b_i< 1$, the linear system \eqref{eq1} has a unique solution, which is given by
\begin{align}\label{eq:b<1}
R_n= \frac{b_n}{(L+t_a)\left(1-\sum_i b_i\right)}, ~~\forall~n.
\end{align}
By this, we have obtained the optimum solution to Problem \eqref{eq_61} for $\sum_i b_i< 1$.

\subsection{Implementation Procedure}
We first express the solutions for $\sum_i b_i\geq 1$ and $\sum_i b_i< 1$ in a unified form.
For $\sum_i b_i< 1$, the solution \eqref{eq:b<1} can also be expressed by \eqref{eq_13},
where $c^*$ and $y^*$ are determined as
\begin{align}\label{eq:c_y}
c^*=1,~ y^*=\frac{1}{(L+t_a)\left(1-\sum_i b_i\right)}.
\end{align}
Our transmission control scheme is described as follows:

$\!\!\!\!\!$\uline{\textbf{\textit{Life-Add}: Lifetime  Adjustable  Design for WiFi Networks}}


\begin{itemize}
\item[(1)] The AP gathers the system parameters $N$, $L$, $t_a$ and $t_s$, and the target energy efficiency parameter $b_i$ from the devices.


\item[(2)] The AP checks if $\sum_i b_i\geq 1$ or $\sum_i b_i< 1$. If $\sum_i b_i\geq 1$, the AP computes $c^*$ and $y^*$ according to \eqref{eq_13} and \eqref{eq_14}; otherwise, it computes $c^*$ and $y^*$ according to \eqref{eq:c_y}.

\item[(3)] The AP broadcasts $c^*$ and $y^*$ to the devices by piggybacking them on its advertisement beacons.

\item[(4)] On hearing $c^*$ and $y^*$, the device $n$ computes $R_n$ by \eqref{eq_13}, and then utilizes $R_n$ to generate its sleep period.
\end{itemize}

\subsection{Performance Analysis}\label{sec:performance_ana}
The performance of Life-Add is characterized by the following theorem:

\textsl{\textbf{Theorem 1}:} Our solution \eqref{eq_13}, \eqref{eq_14}, \eqref{eq_16} and \eqref{eq:c_y} is asymptotically optimal for Problem \textbf{A} as $\frac{t_s}{L+t_a}\rightarrow0$.
%

\begin{proof}
Here, we provide a proof sketch consisting of two steps.
\ifreport
More details of the proof are provided in Appendix \ref{App2}.
\else
We refer to a more detailed proof in \cite{report_scheduling2013}.
\fi


In {Step One}, we show that our solution is asymptotically optimal for Problem \eqref{eq_61} as $\frac{t_s}{L+t_a}\rightarrow0$. If $\sum_i b_i< 1$, we have derived the exact solution to Problem \eqref{eq_61} in Section \ref{sec:sumb<1}. If $\sum_i b_i\geq 1$, the optimum objective value of Problem \eqref{eq_61} is lower bounded by \eqref{eq_15}, which is achieved by our solution \eqref{eq_13}, \eqref{eq_14}, and \eqref{eq_16}. On the other hand, we shown in \cite{report_scheduling2013} that the optimum objective value of Problem \eqref{eq_61} is upper bounded by
\begin{equation}\label{eq26}
N\ln\left(\frac{L}{L+t_a}\right)+\sum_{i} \ln \alpha_i+\sum_i \ln[\min \{b_i,c^*\}].
\end{equation}
Moreover, the gap between the lower bound \eqref{eq_15} and upper bound \eqref{eq26} tends to $0$ as $\frac{t_s}{L+t_a}\rightarrow0$. Therefore, our solution \eqref{eq_13}, \eqref{eq_14}, and \eqref{eq_16} is asymptotically optimal for Problem \eqref{eq_61} as $\frac{t_s}{L+t_a}\rightarrow0$ if $\sum_i b_i\geq 1$.

In {Step Two}, we show that our solution satisfies $R_nt_s\rightarrow0$ as $\frac{t_s}{L+t_a}\rightarrow0$ for both $\sum_i b_i\geq 1$ and $\sum_i b_i< 1$. By this, Problem \eqref{eq_61} and Problem \textbf{A} tends to be the same problem as $\frac{t_s}{L+t_a}\rightarrow0$. Therefore, our asymptotically optimal solution to Problem \eqref{eq_61} is also asymptotically optimal for Problem \textbf{A}.
\end{proof}



\textsl{\textbf{Remark}:} It is worth pointing out that the condition $\frac{t_s}{L+t_a}\approx0$ is satisfied in practical WiFi networks. For example, in IEEE 802.11b WiFi standard, the carrier sensing time is $t_s=4\mu$s~\cite{DBLP:conf/mobicom/MagistrettiCRR11}, and the transmission time of the data packet and ACK varies from 511$\mu$s to 1573$\mu$s for a packet with 0 to 1460 bytes of payload~\cite{ReillyPacketSize}. Hence, $\frac{t_s}{L+t_a}\leq 0.00783$.


%
%

\section{General WiFi Networks with Multiple APs}
\label{sec:multiple}
We now extend Life-Add to more general scenarios with multiple APs, where some clients and/or APs cannot detect the existence of each other because of the large network size. The key challenge here is the complicated interference environment, which makes the transmission control problem very difficult.

\begin{figure}[Ht]
\centering
\includegraphics[width=2in]{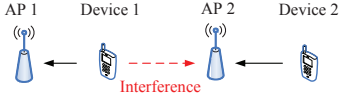}
\caption{An illustrating example of the near-far effect.}
\label{fig:example}
\vspace{-0.5cm}
\end{figure}

One such example is the asymmetric interference scenario shown in Fig. \ref{fig:example}, which is also known as the \emph{near-far effect} in wireless communications. In this example, Device 1 is associated with AP 1 and Device 2 is associated with AP 2. Since AP 1 is out of the interference range of Device 2, it can successfully receive the packets from Device 1. However, the transmissions of Device 2 are seriously interfered by Device 1. In the IEEE 802.11 WiFi designs, Device 1 will keep the minimum backoff since its transmissions are always successful, while Device 2 has high backoff time due to packet collisions at AP 2. This significantly reduces the throughput of Device 2 and results in unfairness between the two devices \cite{hua2008starvation}.  Note that the RTS-CTS mechanism has large overhead, thus it cannot improve the throughput performance significantly.

Another challenge of the multiple AP scenario is the \emph{hidden terminal problem}, where a packet collision may occur at a AP even if the transmission starting time of two packets are more than $t_s$ apart, because the two devices cannot detect each other. Therefore, the collision probability $\beta_{col}$ is under-estimated.

We now extend Life-Add to the multi-AP scenario, and use a heuristic design to handle these two challenges.
In order to compute the $c^{*}$ and $y^*$ in multi-AP networks, each AP overhears the reports of $b_i$ from the devices within its communication range, including the devices that are associated with other APs. Similar overhearing procedure is supported in IEEE 802.11 MAC \cite{IEEE:2007}. For example, to perform virtual carrier sensing, each device overhears the packets from nearby devices to get the packet length information contained in the packet header. In addition, each device receives the beacons broadcasting $c^{*}$ and $y^*$ from all the APs within its communication range.

To resolve the near-far effect issue, Life-Add utilizes the following device collaboration scheme: each device computes the $R_n$s suggested by nearby APs, and chooses to use the smallest $R_n$ value, which corresponds to the largest average sleep duration. This helps to improve fairness among the devices. For example, for the network shown in Fig. \ref{fig:example}, Life-Add will compute the backoff time of Device 1, by considering channel contentions at both AP 1 and AP 2. This is as if nearby APs tell Device 1 how many devices' transmissions it interferes with at each AP (in our example, Device 1 interferes with Device 2 at AP 2). Therefore, Device 1 transmits conservatively to provide more channel access opportunities to Device 2 compared to IEEE 802.11.

For the hidden terminal problem, Life-Add employs the following congestion control scheme similar to that of IEEE 802.11 \cite{IEEE:2007}: Each device increases its average sleep period $1/R'_n$ when it undergoes a timeout, and reduces $1/R'_n$ when it receives an ACK, where $R'_n$ is the actual sleep-wake parameter adopted by the device. The sleep-wake parameter $R_n$ assigned by the APs is used as an upper bound of the actual sleep-wake parameter $R'_n$ to ensure the lifetime constraint.

{Life-Add} for the multiple AP scenario is as follows:

\begin{itemize}
\item[(1)] Each AP gathers the system parameters $L$, $t_a$, $t_s$ and $N$, where $N$ is the number of devices within its communication range (including those associated with other APs). Each AP  also collects $b_i$ from  all these devices.


\item[(2)] Each AP checks whether $\sum_i b_i\geq 1$ or $\sum_i b_i< 1$. If $\sum_i b_i\geq 1$, the AP computes $c^*$ and $y^*$ by  \eqref{eq_13} and \eqref{eq_14}; otherwise, it computes $c^*$ and $y^*$ by \eqref{eq:c_y}.

\item[(3)] Each AP broadcasts $c^*$ and $y^*$ to the devices within its communication range by piggybacking them on its advertisement beacons.

\item[(4)] Each device $n$ may receive multiple pairs of $(c^*,y^*)$ from nearby APs within its communication
range. For each $(c^*,y^*)$, it computes the corresponding $R_n$ using \eqref{eq_13}, and chooses to use the smallest $R_n$ value.

\item[(5)] Each device computes its \emph{Congestion Control Factor} $F_n$. Initially, the value of $F_n$ is 1. Each time the device undergoes a timeout, it doubles its $F_n$ (up to a maximum value of 32). Each time the device receives an ACK, its $F_n$ is reset to 1. The actual sleep-wake parameter $R'_n$ is computed by $R'_n = {R_n}/{F_n}$. Finally, the sleep period is generated as an exponentially distributed random variable with mean $1/R'_n$.
\end{itemize}

Life-Add differs from IEEE 802.11 MAC in the following aspects: (1) In Life-Add, a device transmits immediately after it senses the channel to be idle. However, in IEEE 802.11 MAC, the device keeps sensing the channel for an additional backoff time, which consumes much energy. This is known as the ``idle-listening'' problem in the literature \cite{Zhang:2011:EEI:2030613.2030637}. (2) By using $b_n$ to compute $R_n$, Life-Add takes the energy supply and lifetime requirement of the devices into consideration. (3) Life-Add intends to improve the throughput and fairness performance by carefully controlling the transmission parameters and introducing the device collaboration scheme.

We finally note that a simpler version of Life-Add can be derived by ignoring the lifetime constraints and simply choosing $b_n>1$. This corresponds to the best throughput performance of Life-Add, since the devices do not need
to sacrifice the throughput to satisfy their lifetime constraints. This case will be considered in our simulations in Section \ref{sec:simulation_multi_AP}.

\section{Simulation Results}
\label{sec:simulation}
We now provide  ns-3 \cite{ns3simulator} simulation results to evaluate the performance of Life-Add, where its physical layer is the same as IEEE 802.11b. For the purpose of comparison, we also simulate IEEE 802.11b MAC as well as IEEE 802.11b MAC with RTS-CTS. All the simulation results are averaged over 5-10 system realizations. We evaluate our system under UDP saturation conditions such that the devices always have UDP data to send. The packet length varies according to the measurements in \cite{sinha2007internet}.

\subsection{Single AP Scenario}
We set up a WiFi network with 1 AP and 3 associated devices in a field of size 50m $\times$ 50m. The sensing threshold is set to -100 dBm resulting in a sensing range of approximate 110m. Therefore, all the devices can hear each
other.
The AP is assumed to be connected to the wall power while each device is equipped with a battery and a solar panel that provides renewable energy charging. 
We assume that the devices are HTC Tilt smartphones, where if the WiFi radio is awake, the device consumes\footnote{It was measured in \cite{DBLP:conf/mobisys/RoznerNRR10} that the energy consumption rate of the WiFi radio of HTC Tilt 8900 series in awake mode is 1120 mW and the base energy consumption rate is within 155mW~475 mW (an average of 315 mW) for a total consumption of 1435 mW. When the radio is in sleep mode, the total consumption is 315 mW + 72 mW = 387 mW.} 1435 mW, and if the WiFi radio is in sleep mode, the energy consumption is 387 mW. Each device has a battery with a capacity of 1200 mAh.


\subsubsection{Homogeneous Devices}
\begin{figure}[Ht]
  \centering
    \subfigure[][Average actual lifetime vs. target lifetime]{\label{fig:sim:horizon:time}\includegraphics[width=0.24\textwidth]{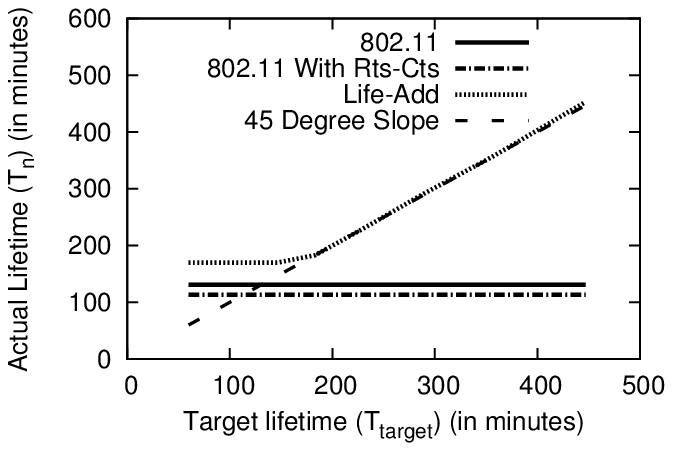}}
  \subfigure[][Total Utility vs. actual lifetime]{\label{fig:sim:horizon:utility}\includegraphics[width=0.24\textwidth]{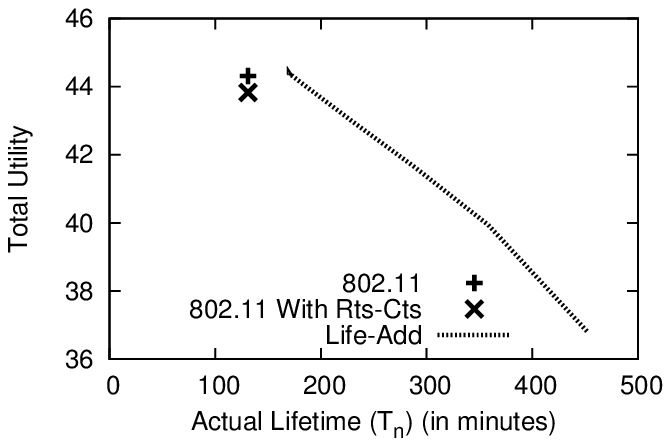}}
  \caption{Simulation results for single AP networks with homogenous devices.}
  \label{fig:utiltput-varyingHorizon}
\end{figure}

We consider a homogeneous scenario where all devices have the same initial battery level (300 mAh) and recharging rate (160 mW).
Fig. \ref{fig:sim:horizon:time} shows the average actual lifetime of the three designs with varying {target lifetime} $T_{target}$. The actual lifetime of Life-Add increases with respect to the target lifetime, while the actual lifetimes of the other two designs stay constant. This figure shows that the actual lifetime of Life-Add is adjustable and satisfies the lifetime constraint \eqref{eq:lifetime_con}. We note that even the minimum lifetime value of Life-Add is longer than that of the IEEE 802.11 MACs, because there is no idle-listening in Life-Add.

Fig. \ref{fig:sim:horizon:utility} shows how Life-Add realizes  lifetime
extension by trading the total utility.
The utility of a device is determined by $U(x)=\ln(x)$, if the average throughput of the device is $x$ kbps.
Life-Add can achieve tunable lifetime by adjusting the average sleeping-backoff durations of the devices. However, for the IEEE 802.11 MACs, it is not possible to adjust the utility and the actual lifetime, and thus, the devices always have fixed utility and lifetime.


\subsubsection{Heterogeneous Devices}
\begin{figure}[Ht]
  \centering
    \label{fig:sim:hetro:util}\includegraphics[width=0.30\textwidth]{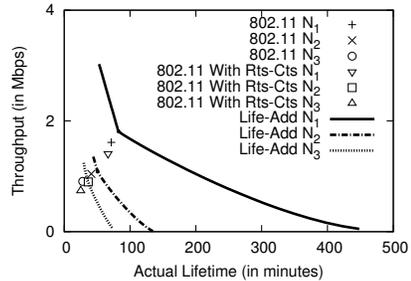}
  \caption{Simulation results for single AP networks with heterogeneous devices.}
  \label{fig:utiltput-hetro}
\vspace{-0.3cm}
\end{figure}

We now explore the performance of Life-Add for a heterogeneous scenario. The initial battery level $B_n$, energy recharging rate $r_n$, and target lifetime values $T_{target,n}$ of the three devices are provided in Table \ref{tab:sim:hetro}, where the varying step sizes of $T_{target,n}$ for the three devices are different. Fig. \ref{fig:utiltput-hetro} shows the tradeoff between the throughput and actual lifetime of all three devices. We can see that Life-Add can achieve higher throughput than the IEEE 802.11 MACs for
each  device, which also implies a larger total utility.


\begin{table} \caption{Parameters of the heterogeneous devices.} \label{tab:sim:hetro} \centering
\begin{tabular}{ l | c | c | c }
    \hline
  Node & Initial Battery & Recharging & Target lifetime\\
& Level (in mAh) & Rate (in mW) &  (in mins.)\\ \hline
    $N_1$ & 200 & 187 & 18,36$,\cdots$,180\\ \hline
    $N_2$ & 100 & 90 & 9,18$,\cdots$,90\\ \hline
    $N_3$ & 66.6 & 67 & 6,12$,\cdots$,60\\
    \hline
\end{tabular}
\vspace{-0.3cm}
\end{table}


\subsection{Multiple AP scenario with Heterogeneous Devices}\label{sec:simulation_multi_AP}
\subsubsection{With Varying Number of Devices}
\label{subsec:hetro:devices:vary}
We setup a field of 300m $\times$ 300m, in which we randomly deploy 4 APs and varying number of devices. Each device is associated with the AP that
has the strongest signal, and the device only sends data packets to its associated AP. The devices have heterogeneous energy supplies. In particular, 1/3 devices
only have access to  the energy stored in their batteries. Another 1/3 devices have access to both battery energy and renewable energy from solar panel. The last 1/3 devices are connected to the wall power that provides high energy recharging rate. The initial battery levels of all the devices are randomly generated by a uniform distribution within the range of $[200,1000]$ mAh. The lifetime constraints are ignored here by choosing $b_i>1$, which achieves the best possible throughput performance.

\begin{figure}[Ht]
  \centering
    \subfigure[][Average actual lifetime vs. number of devices]
    {\label{fig:sim:mc:hetro:nonodes:time}\includegraphics[width=0.24\textwidth]{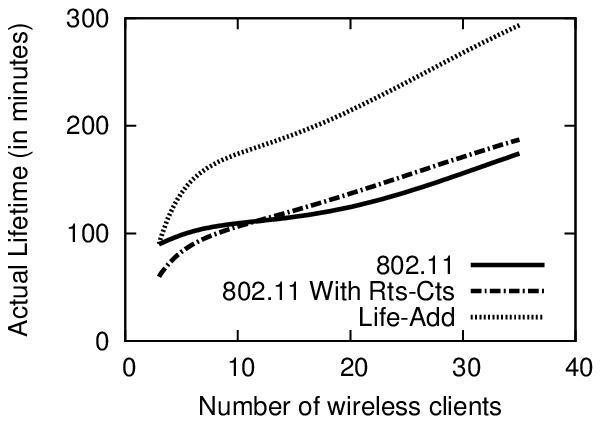}}
  \subfigure[][Total utility vs. number of devices]{\label{fig:sim:mc:hetro:nonodes:util}\includegraphics[width=0.24\textwidth]{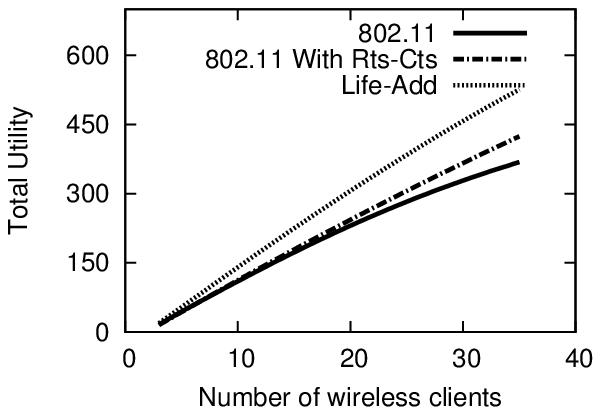}}
  \caption{Simulation results for multi AP networks with heterogeneous devices: varying number of devices.}
  \label{fig:mc:hetro:nonodes:utiltput-varyingNoDevices}
  \vspace{-0.3cm}
\end{figure}

Figure \ref{fig:sim:mc:hetro:nonodes:time} shows the average lifetime performance versus the number of devices.
Life-Add prolongs the average lifetime 
because there is no idle-listening.
Figure \ref{fig:sim:mc:hetro:nonodes:util} shows the total utility provided by these three designs. One can see that Life-Add achieves a larger proportionally fair utility.

\subsubsection{CDF}
\label{subsec:sim:cdf}
\begin{figure}[Ht]
  \centering
    \subfigure[][CDF of the actual lifetime of the devices]{\label{fig:sim:mc:hetro:time}\includegraphics[width=0.24\textwidth]{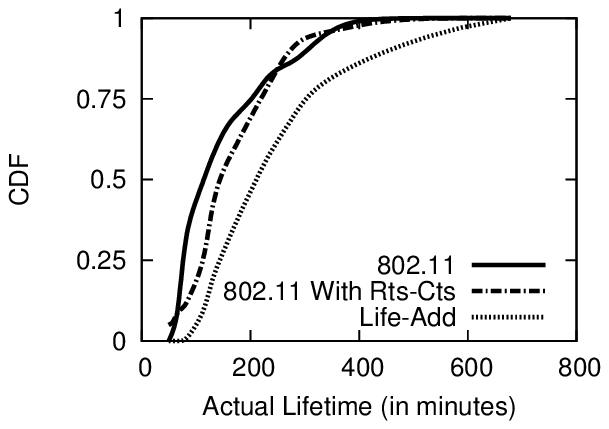}}
  \subfigure[][CDF of the throughput of the devices]{\label{fig:sim:mc:hetro:util}\includegraphics[width=0.24\textwidth]{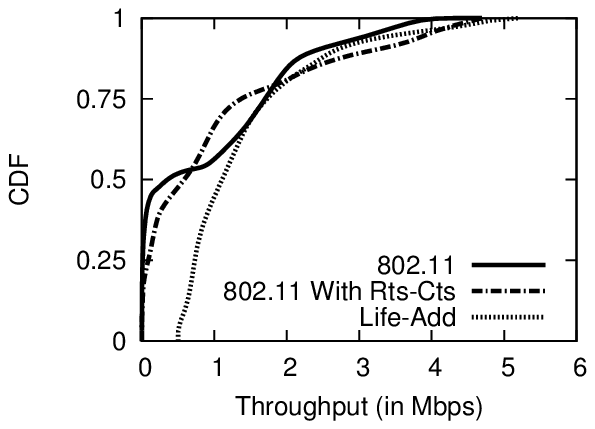}}
  \caption{Simulation results for multi AP networks with heterogeneous devices: CDF performance.}
  \label{fig:mc:hetro:utiltput-varyingNoDevices}
  \vspace{-0.3cm}
\end{figure}

Now we consider a field of 500m $\times$ 500m, in which we randomly deploy 4 APs and 30 devices. The energy supply of the devices are set as in Section \ref{subsec:hetro:devices:vary}.
Figure \ref{fig:sim:mc:hetro:time} shows the CDF (cumulative distribution function) of the actual lifetime of the devices. The figure only plots the CDF for the first 20 devices since the last 10 devices have infinite lifetime due to the high recharging rate provided by the wall power. Figure \ref{fig:sim:mc:hetro:util} shows the CDF of the throughput of the individual devices (for all 30 devices). The average lifetime, average throughput, Jain's fairness index \cite{Jain:1991}, and successful transmission probability for this simulation scenario is provided in Table \ref{tab:compare}. Life-Add has better throughput performance because IEEE 802.11 has too many collisions and the RTS-CTS mechanism is not efficient. It is relative easier to understand the lifetime and fairness improvement. 

\subsubsection{Coexistence with IEEE 802.11}
\begin{figure}[Ht]
  \centering
    \subfigure[][CDF of the actual lifetime of the devices]{\label{fig:sim:mc:coexist:time}\includegraphics[width=0.24\textwidth]{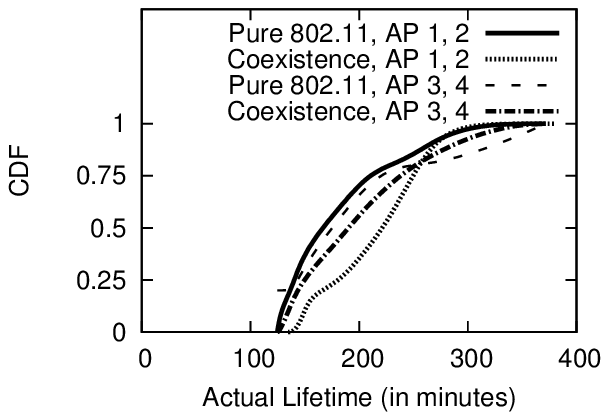}}
  \subfigure[][CDF of the throughput of the devices]{\label{fig:sim:mc:coexist:util}\includegraphics[width=0.24\textwidth]{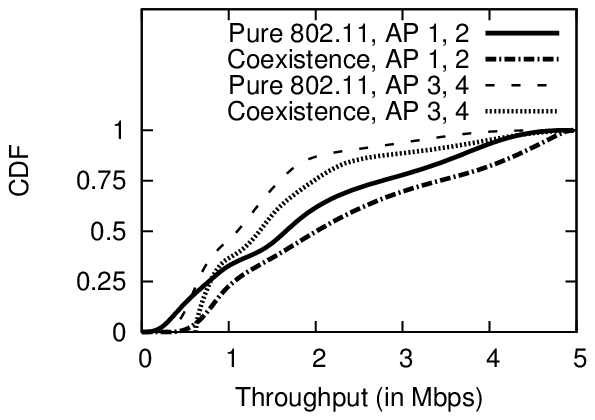}}
  \caption{Simulation results for the coexistence of Life-Add and IEEE 802.11.}
  \label{fig:mc:coexist:utiltput-varyingNoDevices}
  \vspace{-0.3cm}
\end{figure}

Finally, we explore the coexistence between IEEE 802.11 and Life-Add. We set up the network as explained in Section \ref{subsec:sim:cdf}. However, this time AP 1 and 2 (randomly chosen) switch from IEEE 802.11 (without RTS-CTS) to Life-Add while AP 3 and 4 stick to IEEE 802.11. Each device is associated with the AP with the strongest signal, which then determines the design used by the device.
Fig. \ref{fig:mc:coexist:utiltput-varyingNoDevices} shows the lifetime and throughput CDFs of the devices associated with AP 1-2 and AP 3-4, before and after AP 1-2 switch to Life-Add.
One can observe that, after AP 1-2 switch to Life-Add, the lifetime distribution of the devices using Life-Add is significantly improved, because there is no idle-listening in Life-Add. The lifetime distribution of the devices stick to IEEE 802.11 does not change much. On the other hand, the throughput distributions for both groups of devices are improved.
%
%
Therefore, upgrading IEEE 802.11 to Life-Add is beneficial for both the devices switching to Life-Add and those sticking to IEEE 802.11, which provides the incentive to employ Life-Add.


\section{Conclusion}\label{sec_conclusion}
In this paper, we have proposed Life-Add, a life-time adjustable design for WiFi networks. By using a sleep-wake channel contention scheme, Life-Add can resolve the high energy consumption of the current WiFi design. Then, by controlling the sleep period parameters carefully and introducing collaboration among the devices, Life-Add has a small packet collision probability, which implies high throughput. Finally, by providing high priority to the low throughput devices, Life-Add improves the fairness performance. Our simulation results show that Life-Add has better lifetime, throughput, and fairness performance than IEEE 802.11b, and coexists harmoniously with IEEE 802.11b. For future work, we are evaluating how Life-Add behaves under TCP traffic and in the presence of downlink traffic. Its hardware implementation is currently under preparation.

\bibliographystyle{IEEEtran}
\bibliography{citations}

\begin{thebibliography}{10}
\providecommand{\url}[1]{#1}
\csname url@samestyle\endcsname
\providecommand{\newblock}{\relax}
\providecommand{\bibinfo}[2]{#2}
\providecommand{\BIBentrySTDinterwordspacing}{\spaceskip=0pt\relax}
\providecommand{\BIBentryALTinterwordstretchfactor}{4}
\providecommand{\BIBentryALTinterwordspacing}{\spaceskip=\fontdimen2\font plus
\BIBentryALTinterwordstretchfactor\fontdimen3\font minus
  \fontdimen4\font\relax}
\providecommand{\BIBforeignlanguage}[2]{{%
\expandafter\ifx\csname l@#1\endcsname\relax
\typeout{** WARNING: IEEEtran.bst: No hyphenation pattern has been}%
\typeout{** loaded for the language `#1'. Using the pattern for}%
\typeout{** the default language instead.}%
\else
\language=\csname l@#1\endcsname
\fi
#2}}
\providecommand{\BIBdecl}{\relax}
\BIBdecl

\bibitem{Zhang:2011:EEI:2030613.2030637}
X.~Zhang and K.~G. Shin, ``{E-MiLi}: Energy-minimizing idle listening in
  wireless networks,'' in \emph{MobiCom}, 2011, pp. 205--216.

\bibitem{DBLP:conf/mobisys/RoznerNRR10}
E.~Rozner, V.~Navda, R.~Ramjee, and S.~K. Rayanchu, ``{NAPman}:
  Network-assisted power management for {WiFi} devices,'' in \emph{MobiSys},
  2010, pp. 91--106.

\bibitem{Baiamonte:2006:SED:1147589.1147605}
V.~Baiamonte and C.-F. Chiasserini, ``Saving energy during channel contention
  in 802.11 {WLANs},'' \emph{Mob. Netw. Appl.}, vol.~11, no.~2, pp. 287--296,
  2006.

\bibitem{SolarMobileCharger1}
{Firenew Technology Company}, ``Solar mobile charger,''
  \url{http://www.firenew.com.cn/cn/product.asp?nclassid=248&typeid=71}.

\bibitem{portable_batteries}
{Mophie Inc.}, ``External {USB} portable battery charger,''
  \url{http://www.mophie.com/category-s/53.htm}.

\bibitem{Jain:1991}
R.~Jain, \emph{The Art of Computer Systems Performance Analysis}.\hskip 1em
  plus 0.5em minus 0.4em\relax John Wiley \& Sons, 1991.

\bibitem{840210}
G.~Bianchi, ``Performance analysis of the {IEEE} 802.11 distributed
  coordination function,'' \emph{IEEE J. Sel. Areas Commun.}, vol.~18, no.~3,
  pp. 535--547, Mar. 2000.

\bibitem{5340575}
L.~Jiang and J.~Walrand, ``A distributed {CSMA} algorithm for throughput and
  utility maximization in wireless networks,'' \emph{IEEE/ACM Trans. Netw.},
  vol.~18, no.~3, pp. 960--972, Jun. 2010.

\bibitem{Neely:2006:EOC:2263442.2272284}
M.~J. Neely, ``Energy optimal control for time-varying wireless networks,''
  \emph{IEEE Trans. Inf. Theory}, vol.~52, no.~7, pp. 2915--2934, 2006.

\bibitem{Lin:2010:LDE:1816262.1816275}
L.~Lin, X.~Lin, and N.~B. Shroff, ``Low-complexity and distributed energy
  minimization in multihop wireless networks,'' \emph{IEEE/ACM Trans. Netw.},
  vol.~18, no.~2, pp. 501--514, 2010.

\bibitem{6195775}
S.~Chen, P.~Sinha, N.~Shroff, and C.~Joo, ``A simple asymptotically optimal
  energy allocation and routing scheme in rechargeable sensor networks,'' in
  \emph{IEEE INFOCOM 2012}, Mar. 2012, pp. 379--387.

\bibitem{Huang:2011:UOS:2107502.2107531}
L.~Huang and M.~J. Neely, ``Utility optimal scheduling in energy harvesting
  networks,'' in \emph{MobiHoc}, 2011.

\bibitem{5339116}
J.~Kim, X.~Lin, N.~Shroff, and P.~Sinha, ``Minimizing delay and maximizing
  lifetime for wireless sensor networks with anycast,'' \emph{IEEE/ACM Trans.
  Netw.}, vol.~18, no.~2, pp. 515--528, Apr. 2010.

\bibitem{Zhao:2012:SPM:2379776.2379783}
Y.~Z. Zhao, C.~Miao, M.~Ma, J.~B. Zhang, and C.~Leung, ``A survey and
  projection on medium access control protocols for wireless sensor networks,''
  \emph{ACM Comput. Surv.}, vol.~45, no.~1, pp. 7:1--7:37, 2012.

\bibitem{5451759}
A.~Bachir, M.~Dohler, T.~Watteyne, and K.~Leung, ``{MAC} essentials for
  wireless sensor networks,'' \emph{IEEE Commun. Surv. Tutorials}, vol.~12,
  no.~2, pp. 222--248, Quarter 2010.

\bibitem{DBLP:journals/winet/AnastasiCGP08}
G.~Anastasi, M.~Conti, E.~Gregori, and A.~Passarella, ``802.11 power-saving
  mode for mobile computing in {Wi-Fi} hotspots: Limitations, enhancements and
  open issues,'' \emph{Wireless Networks}, vol.~14, no.~6, pp. 745--768, 2008.

\bibitem{Gallager96}
R.~G. Gallager, \emph{{Discrete Stochastic Processes}}.\hskip 1em plus 0.5em
  minus 0.4em\relax Boston: Kluwer Academic Publishers, 1996.

\bibitem{Kelly97chargingand}
F.~Kelly, ``Charging and rate control for elastic traffic,'' \emph{European
  Transactions on Telecommunications}, 1997.

\bibitem{report_scheduling2013}
\BIBentryALTinterwordspacing
S.~Chen, T.~Bansal, Y.~Sun, P.~Sinha, and N.~B. Shroff, ``{Life-Add}: Lifetime
  {Adjustable} design for {WiFi} networks with heterogeneous energy supplies,''
  2013, {Technical Report, Depts of ECE and CSE, Ohio State University}.
  [Online]. Available: \url{http://www.ece.osu.edu/~chens/wiopt13.pdf}
\BIBentrySTDinterwordspacing

\bibitem{DBLP:conf/mobicom/MagistrettiCRR11}
E.~Magistretti, K.~K. Chintalapudi, B.~Radunovic, and R.~Ramjee, ``{WiFi-Nano}:
  reclaiming {WiFi} efficiency through 800 ns slots,'' in \emph{MOBICOM}, 2011.

\bibitem{ReillyPacketSize}
M.~Gast, ``{When Is 54 Not Equal to 54? A Look at 802.11a, b, and g
  Throughput},'' \url{http://www.oreillynet.com/pub/a/wireless/2003
  /08/08/wireless\_throughput.html?page=1}.

\bibitem{hua2008starvation}
C.~Hua and R.~Zheng, ``Starvation modeling and identification in dense 802.11
  wireless community networks,'' in \emph{IEEE INFOCOM 2008}, 2008, pp.
  1022--1030.

\bibitem{IEEE:2007}
\emph{{Wireless LAN Medium Access Control (MAC) and Physical Layer (PHY)
  specifications}}, Institute of Electrical and Electronics Engineers, 2007,
  http://standards.ieee.org/about/get/802/802.11.html.

\bibitem{ns3simulator}
``{ns-3},'' \url{http://www.nsnam.org/}.

\bibitem{sinha2007internet}
R.~Sinha, C.~Papadopoulos, and J.~Heidemann, ``Internet packet size
  distributions: Some observations,'' \emph{USC/Information Sciences
  Inst.[Online]. Available: http://netweb. usc. edu/\~{} rsinha/pkt-sizes},
  2007.

\bibitem{Boyd:2004:CO:993483}
S.~Boyd and L.~Vandenberghe, \emph{Convex Optimization}.\hskip 1em plus 0.5em
  minus 0.4em\relax New York, NY, USA: Cambridge University Press, 2004.

\end{thebibliography}
\ifreport

\appendices

\section{Derivation of \eqref{eq_4_3}}\label{App00}
Let $\gamma_n$ denote the probability that node $n$ transmits in a sleep-wake cycle, no matter whether traffic collision occurs. According to our sleep-wake scheduling scheme, node $n$ transmits as long as no other node wakes up before $X_n-t_s$, i.e., $X_i\geq X_n-t_s$ for all $i\in \mathcal{N}$ and $i\neq n$.
The probability $\gamma_n$ is given by
\begin{eqnarray}
&&\!\!\!\!\!\!\!\gamma_n= \Pr(X_i\geq X_n-t_s  \forall i\neq n)\nonumber\\
&&\!\!\!\!\!\!\!~~~~=\Pr(X_i\geq X_n-t_s,  \forall i\neq n, X_n\geq t_s)+\Pr(X_n<t_s), \nonumber
\end{eqnarray}
where the first term in the RHS is determined by
\begin{align}
&~~~\Pr(X_i\geq X_n-t_s\geq0,\forall i\neq n)\nonumber\\
&\overset{}{=} \mathbb{E}[\Pr(X_i\geq X_n-t_s\geq0, \forall i\neq n|X_n)]\nonumber\\
&\overset{}{=} \mathbb{E}\left[\prod_{i\neq n}\Pr(X_i\geq X_n-t_s\geq0|X_n)\right]\nonumber\\
&=\int_{t_s}^{\infty}\!\left[\prod_{i\neq n}\exp(-R_i (x_n-t_s))\right]\!R_n\exp(-R_{n}x_n)dx_n\nonumber\\
&=\exp(-R_nt_s)\frac{R_n}{\sum_{i}R_i},\nonumber
\end{align}
and the first term in the RHS is given by
\begin{align}
&\Pr(X_n<t_s)=\int_{0}^{t_s}f(x_n)dx_n\nonumber\\
&=1-\exp(-R_nt_s).\nonumber
\end{align}
Thus, we have
\begin{align}
\gamma_n=1-\exp(-R_nt_s)+\exp(-R_nt_s)\frac{R_n}{\sum_{i}R_i}.\nonumber
\end{align}

Similar with \eqref{eq_4_2}, we can obtain that
\begin{align}
\!\!\!P_n&=\frac{\gamma_n \left(\mathbb{E}[\mathbf{trans}]+\mathbb{L}[\mathbf{ACK}]\right)}{(\sum_i\beta_i +\beta_{col}) (\mathbb{E}[\mathbf{trans}]+\mathbb{L}[\mathbf{ACK}])+\mathbb{E}[\mathbf{Idle}]}\nonumber\\
&=\frac{\gamma_n(L+t_a)}{L+t_a+\frac{1}{\sum_{i}R_i}}\nonumber\\
&=\frac{[1-\exp(-R_nt_s)]\sum_{i}R_i+\exp(-R_nt_s)R_n}{\sum_i R_i+\frac{1}{L+t_a}}.\nonumber
\end{align}

\section{Proof of \eqref{eq_13} and \eqref{eq_14}} \label{App1}
Let $\tau_n$ and $\nu$ be the Lagrange multiplier associated with the constraints \eqref{eq_41} and \eqref{eq_42}.
According to the Karush-Kuhn-Tucker (KKT) necessary conditions \cite{Boyd:2004:CO:993483}, we know that the optimum solution to \eqref{eq_12} must satisfy
\begin{eqnarray}
&&-\frac{1}{R_n}+\nu +\tau_n=0,\\
&&\tau_n\geq 0,R_n- b_n y\leq 0,\\
&&\tau_n(R_n- b_n y)=0,\\
&&\sum_i R_i=y.
\end{eqnarray}
If $\tau_n=0$, then $R_n = \frac{1}{\nu}$ and $R_n- b_n y\leq 0$; otherwise, if $\tau_n>0$, then $R_n= b_n y$ and $R_n< \frac{1}{\nu}$. Therefore,
\begin{eqnarray}
R_n = \min\left\{b_n y,\frac{1}{\nu^*}\right\},
\end{eqnarray}
where $\nu^*$ satisfies the constraint \eqref{eq_42}.
Define $c^*=\frac{1}{\nu^* y}$. If $\sum_i b_i\geq 1$, one can see that the optimum solution to \eqref{eq_12} is given by \eqref{eq_13} and \eqref{eq_14}.

\section{Proof of \eqref{eq1}}\label{App0}
Let $\eta_n$ be the Lagrange multiplier associated with the constraints of \eqref{eq_B0}. According to the KKT necessary conditions \cite{Boyd:2004:CO:993483}, we know that the optimum solution to \eqref{eq_B0} must satisfy
\begin{eqnarray}\label{eq:app1}
&&-(N-1)t_s+\eta_n(1-b_n)=0,\\
&&\eta_n\geq 0,R_n- b_n\left(\sum_i R_i+\frac{1}{L+t_a}\right)\leq 0,\\
&&\eta_n\left[R_n- b_n\left(\sum_i R_i+\frac{1}{L+t_a}\right)\right]=0.\label{eq:app2}
\end{eqnarray}
By \eqref{eq:app1} and $\sum_i b_i< 1$, we have $\eta_n>0$. Then, in view of \eqref{eq:app2}, the optimum solution to \eqref{eq_B0} satisfies \eqref{eq1}.
%
%

\section{Proof of Theorem 1}\label{App2}

\subsection{Step One}
For $\sum_i b_i< 1$, It is known that \eqref{eq:b<1} is optimum for Problem \eqref{eq_61}. Therefore, we only need to consider the case of
$\sum_i b_i\geq 1$. For this case, we will show the optimum value of Problem \eqref{eq_61} is lower bounded by \eqref{eq_15}, i.e., the optimum value achieved by our solution \eqref{eq_13}, \eqref{eq_14}, and \eqref{eq_16}, and upper bounded by
\begin{equation}\label{eq26}
N\ln\left(\frac{L}{L+t_a}\right)+\sum_{i} \ln \alpha_i+\sum_i \ln[\min \{b_i,c^*\}],
\end{equation}
and the gap between the upper and lower bounds tends to $0$ as $\frac{t_s}{L+t_a}\rightarrow0$.

Let $f^*$ denote the optimum value of Problem \eqref{eq_61}. The energy constraint in \eqref{eq_41} is more stringent than \eqref{eq_6}, since the term $\frac{1}{L+t_a}$ is omitted. Therefore, Problem \eqref{eq_12} has a smaller feasible region than Problem \eqref{eq_61}, its optimum solution \eqref{eq_13}-\eqref{eq_16} is feasible for Problem \eqref{eq_61}. By this, our solution \eqref{eq_13}-\eqref{eq_16} provides a lower bound of $f^*$, which is given by \eqref{eq_15}.

Now we consider the upper bound of $f^*$.
Define an auxiliary variable $z=\sum_n R_n+\frac{1}{L+t_a}$. Given $z$, Problem \eqref{eq_61} can be reformulated as the following problem:
\begin{align}\label{eq62}
& \max\limits_{R_n>0} \sum_{n} \ln R_n-N\ln z+\sum_{n} \ln \alpha_n+N\ln\left(\frac{L}{L+t_a}\right)\nonumber\\
& ~~~~~~~~~~~~-\!(N\!-\!1)\left(z\!-\!\frac{1}{L+t_a}\right)t_s,\\
& ~~\textrm{s.t.}~  R_n\leq b_n z, \forall n \label{eq22}\\
& ~~~~~~ \sum_n R_n+\frac{1}{L+t_a}=z.\label{eq21}
\end{align}
Let $\lambda_n$ and $\mu$ be the Lagrange multiplier associated with the constraints \eqref{eq22} and \eqref{eq21}. Since \textrm{Problem} \eqref{eq62} is a convex optimization problem, according to the KKT conditions, the optimum solution of \textrm{Problem} \eqref{eq62} satisfies
\begin{eqnarray}
&&-\frac{1}{R_n}+\mu +\lambda_n=0,\nonumber\\
&&\lambda_n\geq 0,R_n- b_n z\leq 0,\nonumber\\
&&\lambda_n(R_n- b_n z)=0,\nonumber\\
&&\sum_n R_n+\frac{1}{L+t_a}=z.\nonumber
\end{eqnarray}
If $\lambda_n=0$, then $R_n = \frac{1}{\mu}$ and $R_n- b_n z\leq 0$; otherwise, if $\lambda_n>0$, then $R_n= b_n z$ and $R_n< \frac{1}{\mu}$. Therefore,
\begin{eqnarray}
R_n = \min\left\{b_n z,\frac{1}{\mu^*}\right\},\nonumber
\end{eqnarray}
where $\mu^*$ satisfies the constraint \eqref{eq21}.
Define $c^*(z)=\frac{1}{\mu^* z}$, the optimum solution to \textrm{Problem} \eqref{eq62} is given by
\begin{eqnarray}\label{eq23}
R_n = \min\{b_n,c^*(z)\}z,
\end{eqnarray}
where $c^*(z)>0$ satisfies the constraint
\begin{eqnarray}\label{eq24}
\sum_n\min\{b_n,c^*(z)\}z+\frac{1}{L+t_a}=z.
\end{eqnarray}
Substituting \eqref{eq23} and \eqref{eq24} into \textrm{Problem} \eqref{eq62}, we get
\begin{eqnarray}
&\!\!\!\!\!f^* =\max\limits_{z>\frac{1}{L+t_a}}\sum_n \ln\min\{b_n,c^*(z)\}-\!(N\!-\!1)(z\!-\!\frac{1}{L+t_a})t_s\!\!\!\!\!\!\nonumber\\
&\hspace{1.8cm}+N\ln\left(\frac{L}{L+t_a}\right)+\sum_{n} \ln \alpha_n\nonumber\\
&  \!\!\!\!\!\!\textrm{s.t.}~~ \sum_n\min\{b_n,c^*(z)\}z+\frac{1}{L+t_a}=z.\nonumber
\end{eqnarray}
By omitting the second negative term in this problem, we drive that $f^*$ is upper bounded by the optimum value of the following problem:
\begin{eqnarray}\label{eq63}
&\!\!\!\!\!\!\!\!\!\!\!\!\!\!\!\!\!\!\!\!\!\!\!\!\!\!\!\!\!\!\max\limits_{z>\frac{1}{L+t_a}}\sum_n \ln\min\{b_n,c^*(z)\}\\
&\hspace{1.8cm}+N\ln\left(\frac{L}{L+t_a}\right)+\sum_{n} \ln \alpha_n\nonumber\\
&  ~\textrm{s.t.}~~ \sum_n\min\{b_n,c^*(z)\}z+\frac{1}{L+t_a}=z.\label{eq25}
\end{eqnarray}
The objective function of \eqref{eq63} is increasing in $c^*(z)$. Moreover,
according to the constraint \eqref{eq25}, $c^*(z)$ is strictly increasing in $z$, if $\sum_i b_i\geq 1$. Therefore, the optimum value of $z$ to Problem \eqref{eq63} is $z=\infty$. At the limit, the constraint \eqref{eq25} reduces to \eqref{eq_14} and $c^*(z)=c^*$. Therefore, $f^*$ is upper bounded by \eqref{eq26}.

The gap between the upper bound \eqref{eq26}
and lower bound \eqref{eq_15} is given by the optimum value of the following problem:
\begin{eqnarray}\label{eq27}
\max_{y>0} N \ln y-N\ln\left(y+\frac{1}{L+t_a}\right)-(N-1)yt_s,\nonumber
\end{eqnarray}
where the optimum solution $y^*$ is given by \eqref{eq_16}. According to \eqref{eq_16}, we can see that
\begin{eqnarray}
&&~~~\lim_{\frac{t_s}{L+t_a}\rightarrow0} \frac{1}{y^*(L+t_a)}\nonumber\\
&&=\lim_{\frac{t_s}{L+t_a}\rightarrow0}\frac{2}{-1+\sqrt{1+\frac{4N(L+t_a)}{(N-1)t_s}}}=0.\nonumber
\end{eqnarray}
Therefore, we have
\begin{eqnarray}
&&~~~\lim_{\frac{t_s}{L+t_a}\rightarrow0} \ln\left(\frac{y^*}{y^*+\frac{1}{L+t_a}}\right)\nonumber\\
&&=\lim_{\frac{t_s}{L+t_a}\rightarrow0} \ln \left(\frac{1}{1+\frac{1}{y^*(L+t_a)}}\right)=0.\nonumber
\end{eqnarray}
On the other hand, we also have
\begin{eqnarray}\label{eq64}
&&~~~\lim_{\frac{t_s}{L+t_a}\rightarrow0} y^*t_s \nonumber\\
&&=\lim_{\frac{t_s}{L+t_a}\rightarrow0}-\frac{t_s}{2(L+t_a)}\nonumber\\
&&~~~+\sqrt{\frac{t_s^2}{4(L+t_a)^2}+\frac{Nt_s}{(N-1)(L+t_a)}}\nonumber\\
&&=0.
\end{eqnarray}
Therefore, the gap between the upper and lower bounds tends to $0$ as $\frac{t_s}{L+t_a}\rightarrow0$.

\subsection{Step Two}
For both the cases of $\sum_i b_i\geq 1$ and $\sum_i b_i< 1$, it is not hard to show that $R_nt_s\rightarrow0$ as $\frac{t_s}{L+t_a}\rightarrow0$. Hence, $[1-\exp(-R_nt_s)]\sum_{i}R_i+\exp(-R_nt_s)R_n \rightarrow R_n$ as $\frac{t_s}{L+t_a}\rightarrow0$. By this, Problem \eqref{eq_61} and Problem \textbf{A} tends to be the same problem as $\frac{t_s}{L+t_a}\rightarrow0$. Therefore, our asymptotically optimal solution to Problem \eqref{eq_61} is also asymptotically optimal for Problem \textbf{A}. By this, the asserted result is proved.

\else
\fi

\end{document}